\documentclass[%
 reprint,
 amsmath,amssymb,
 aps,
]{revtex4-2}

\usepackage{graphicx}
\usepackage{dcolumn}
\usepackage{bm}
\usepackage[colorlinks=true,linkcolor=blue,citecolor=blue,urlcolor=blue]{hyperref}

\allowdisplaybreaks

\begin{document}

\preprint{APS/123-QED}

\title{K-shell ionization and characteristic x-ray radiation by high-energy electrons and positrons in oriented silicon crystals}

\author{S.V. Trofymenko}
\email{trofymenko@kipt.kharkov.ua}
\author{I.V. Kyryllin}
\email{i.kyryllin@gmail.com}
\affiliation{Akhiezer Institute for Theoretical Physics of the National Science Center ``Kharkiv Institute of Physics and Technology'', 1 Akademichna st., 61108 Kharkiv, Ukraine}
\affiliation{V.N. Karazin Kharkiv National University, 4 Svobody sq., 61022 Kharkiv, Ukraine}


\date{\today}

\begin{abstract}
K-shell ionization and characteristic x-ray radiation (CXR) by high-energy electrons and positrons in oriented silicon crystals are studied using computer simulation. A method for this simulation has been developed and is described in detail. The evolution of the angular distribution of CXR from the upstream surface of the crystal with changes in the angle between the incident particle momentum and the crystal  $\langle100\rangle$ axis or (100) plane, as well as with changes in particle energy over a wide range (1--1000 GeV), is investigated. It is shown that in most cases this evolution is non-monotonic. The physical mechanisms underlying this behavior are discussed. In particular, the impact of the dechanneling process on CXR produced by electrons is analyzed.

\end{abstract}

\pacs{Valid PACS appear here}
\maketitle


\section{Introduction}
\label{Introduction}

Most of the energy that ultrarelativistic electrons lose while moving in a medium is emitted in the form of bremsstrahlung gamma quanta, which propagate nearly parallel to the particle's velocity. \cite{AkhiezerShulgaBook}. A smaller part of the electron's energy is transferred to atomic electrons, leading to excitation and ionization of the atoms, and constitutes the particle's ionization energy loss \cite{SigmundBook}. This component can be separated from bremsstrahlung and measured experimentally. It is possible since, in thin targets, the bremsstrahlung photons predominantly leave the target without depositing their energy inside it, whereas a significant fraction of the energy lost to atomic excitation and ionization can remain within the target (e.g., as electron-hole pairs created in semiconductor detectors \cite{DetectorsBook}). A small portion of the energy lost by the incident particle to atomic ionization in a thin boundary layer of the target leaves it in the form of characteristic radiation and Auger electrons, emitted when vacancies in the ionized atoms are filled by electrons from higher energy levels. It contains valuable information about various physical aspects of atomic ionization by incident particles. This is particularly relevant for characteristic x-ray radiation (CXR) emitted as a result of ionization of atomic K shells. It is relatively weakly absorbed by the medium and can escape from a much thicker boundary layer than the Auger electrons and characteristic radiation from other shells with lower binding energies (L, M, etc.). Although the cross section of K-shell ionization by high-energy incident electrons is the lowest among all shells, its CXR photon yield can be the highest, owing to both the higher probability (fluorescence yield) of radiative transitions during the filling of the shell vacancy and the lower attenuation in the medium, which becomes a significant factor if the target is not ultrathin. Furthermore, CXR from high-energy particles is rather convenient to detect due to its broad angular distribution, allowing the detector to be placed far from the direction of bremsstrahlung, x-ray transition radiation at the particle's entrance into and exit from the target \cite{GinzburgTsytovichBook} and possible background radiation accompanying the particle beam. All these facts make the CXR resulting from K-shell ionization (as well as from L-shell ionization in high-Z media) a convenient tool for studying the structure of inner atomic shells and the processes of interaction of incident particles with them. They also enable practical applications of CXR as a source of monochromatic x-ray photons \cite{Bak1983,TrofymenkoPhysRev2020}, for spectroscopic analysis of materials \cite{MullerXrayAnalysis} and for diagnostics of parameters of high-energy particle beams \cite{Chaikovska,Nazhmudinov2018}. 

A series of studies has been devoted to precise measurements of the cross section $\sigma_K$ of K-shell ionization by ultrarelativistic electrons \cite{Middleman,Dangerfield,Ishii,Hoffmann,Kamiya,Genz,Bak1983,Bak1986,Meyerhof,Spooner,Shchagin1994,Nazhmudinov2021}. Special attention was given to the dependence of $\sigma_K$ on the atomic number of the target and the energy of the incident electrons. In the works \cite{Kolbenstvedt1967,Kolbenstvedt1975,Komarov1980,Amundsen1981,Bak1986,Sorensen1986,Ermilova,Sorensen,Chechin}, theoretical approaches to calculation of $\sigma_K$ and ionization probabilities were developed, and the analysis of some of the cited experiments presented. In particular, it was demonstrated that at high electron energies, in the boundary layer of the target, $\sigma_K$ is significantly influenced by x-ray transition radiation (TR) emitted as the electron enters the target. This leads to the absence of the density effect for $\sigma_K$ in this region and its logarithmic increase with the electron energy $E$. The density effect, predicted by Fermi \cite{Fermi} and thoroughly investigated by Sternheimer et al. (see \cite{Sternheimer} and refs. therein) for the average ionization loss of high-energy particles, leads to the saturation of $\sigma_K$ at a constant value as a function of $E$ and is manifested only deep inside the target (in particular, near its downstream surface if the target is thick enough \cite{Meyerhof,Spooner}).
In \cite{TrofymenkoPhysRev2020}, it was shown that the evolution of the electromagnetic field of a high-energy electron in a periodic multifoil target can enhance $\sigma_K$ (for K-shell ionization in the target foils) by several times compared to its value in the absence of the density effect in a single foil. 

The study of electrodynamic processes occurring during the motion of high-energy charged particles in oriented crystals is of particular interest. When a particle enters a crystal at a small angle~\footnote{
By small angles here we mean angles on the order of the critical angle for axial channeling \cite{Lindhardt1965}. For example, this angle is approximately 385 $\mu$rad for electrons or positrons with an energy of 1 GeV moving in a crystal near the $\langle100\rangle$ axis. At the same time, for electrons or positrons with an energy of 100 GeV, this angle is approximately 38.5 $\mu$rad, since it is inversely proportional to the square root of the particle energy.} with respect to a crystal axis or plane, its motion significantly differs from that in an amorphous medium or a non-oriented crystal. In particular, the particle can be captured in a potential well formed either by individual (for negatively charged particles) or by neighboring (for positively charged ones) atomic planes or strings, and move along these planes or strings, performing small-amplitude oscillations in the transverse direction. This regime of motion is known as channeling \cite{Lindhardt1965}. Peculiarities of particle motion in oriented crystals result in a significant modification of the properties of radiation (see, e.g., \cite{AkhiezerShulgaBook,BaierBook,Uggerhoj2005}) and atomic ionization produced by the particle. In particular, in \cite{Fich1976,Esbensen1977,Esbensen1978,Adejshvili1983,PapeMoller2001}, it was experimentally demonstrated that the probability distribution function of the ionization loss (Landau distribution \cite{Landau1944}) changes when the crystalline detector is oriented along a planar or axial direction relative to the incident particle beam. In \cite{TrofymenkoKyryllin2024,TrofymenkoKyryllin2025}, the evolution of this distribution with the change of the crystal orientation has been investigated using computer simulations.  

Ionization of inner atomic shells by incident charged particles in oriented crystals has been thoroughly investigated experimentally for non-relativistic protons \cite{Brandt1965,Khan1966,Kosse1983,Shchagin1989} and electrons \cite{Hall1966,Grachev1966}. A pronounced reduction in the value of $\sigma_K$ was registered for the channeled protons. It is caused by the fact that protons, being positively charged particles, are repelled by the atoms. In the channeling mode, they do not approach the atoms very closely, which reduces the probability of colliding with a K-shell electron and creating a vacancy in this shell. For electrons, the opposite effect of an increase in CXR yield in oriented crystals was observed, but the effect was noticeably smaller than for protons. For relativistic and ultrarelativistic particles, there are fewer experimental investigations in this field. In all of them, the crystal was oriented near the axial direction relative to the particle beam. In \cite{Bak1982}, a pronounced dip in the orientational dependence of CXR yield from Ge crystal was observed for relativistic protons and $\pi^+$ mesons, when the crystal axis $\langle110\rangle$ was aligned with the particle beam. For $\pi^-$ mesons a smaller peak in this dependence was observed in this case. In \cite{Aleinik1987}, a peak in the orientational dependence of CXR yield from Ge crystal was also observed for 900 MeV electrons when the crystal axis $\langle100\rangle$ was aligned with the particle beam. In \cite{Andreyashkin2001}, such a peak was reported for 600, 800 and 1000 MeV electrons with the $\langle111\rangle$ orientation of a W crystal.


These works demonstrated that a thorough study of K-shell ionization and CXR by high-energy particles in oriented crystals is of interest both from the fundamental and practical points of view. On the one hand, when moving in a channeling mode, the particles are redistributed in the space of impact parameters with respect to atomic planes and strings. This allows probing experimentally the impact parameter dependence of the K-shell ionization probability. On the other hand, measuring the orientational dependence of CXR could serve as a non-destructive method for determining the optimal orientation of a crystal for various experiments and charged particle beam steering, in particular using bent crystals \cite{Sytov2022,Kyryllin2023}.


Theoretical calculation of K-shell ionization cross sections for ultrarelativistic channeled electrons and positrons was carried out in \cite{Darbinian1982}. However, these calculations were analytical and were not based on realistic particle trajectories in the crystal. The authors also did not take into account the gradual onset of the density effect in K-shell ionization as the particle penetrates deeper into the target, which led to an underestimation of the relative magnitude of the crystal orientation effect at high particle energies. In the present work, we develop a method of computer simulation of the process of K-shell ionization by high-energy particles in oriented crystals. The method is based on simulation of particle trajectories in the crystal through numerical solution of the equation of motion in the realistic potential of crystalline planes and atomic strings. We also take into account the evolution of the electromagnetic field around the particle as it enters the crystal, associated with the formation of TR, which leads to the onset of the density effect in K-shell ionization. Such an approach makes it possible to describe the CXR yield from the upstream surface of the target (as well as from the downstream one) as registered by a detector placed at an arbitrary position, in particular at relatively small angles to the direction opposite to the particle motion. In these directions the CXR originates from a boundary layer that is wider than the distance over which the onset of the density effect is almost completed. It also allows taking into account the influence of dechanneling on the ionization and CXR yield, which is significant for electrons. It also allows one to account for the beam angular divergence (however, we neglect it in the present study). 


Using the developed method, we study the evolution of the CXR yield from the upstream surface of a silicon crystal with the change of the angle between the direction of the incident particle beam and the $\langle100\rangle$ axis or the $(100)$ plane. For comparison, some results are also presented for the (110) plane. The consideration is performed for electrons and positrons in a wide range of energies: from 1 GeV to 1 TeV. A peculiar interplay between the effects of dechanneling and TR on the CXR yield is shown to manifest itself in the energy dependence of this yield for electrons.


\section{Simulation method}
\label{sec2} 

In this section, we present a detailed description of the method developed for simulating K-shell ionization and CXR emission by high-energy particles in oriented crystals. In this method, we partially employ the approach developed earlier for simulating ionization energy loss of relativistic particles in crystals \cite{TrofymenkoKyryllin2024,TrofymenkoKyryllin2025}, which demonstrated a quite good agreement with experimental results. We combine it with the approach developed in \cite{Bak1986,Sorensen} for calculating K-shell ionization cross sections in amorphous targets, which is based on the application of the equivalent photon method for distant collisions of the impinging particle with atomic electrons and the binary collision approximation for close collisions.

\subsection{Contribution from the region with $\rho<\rho_0$}
\label{subsec_p<p0}

\begin{figure}
	\begin{centering}
		\includegraphics[width=0.4\textwidth]{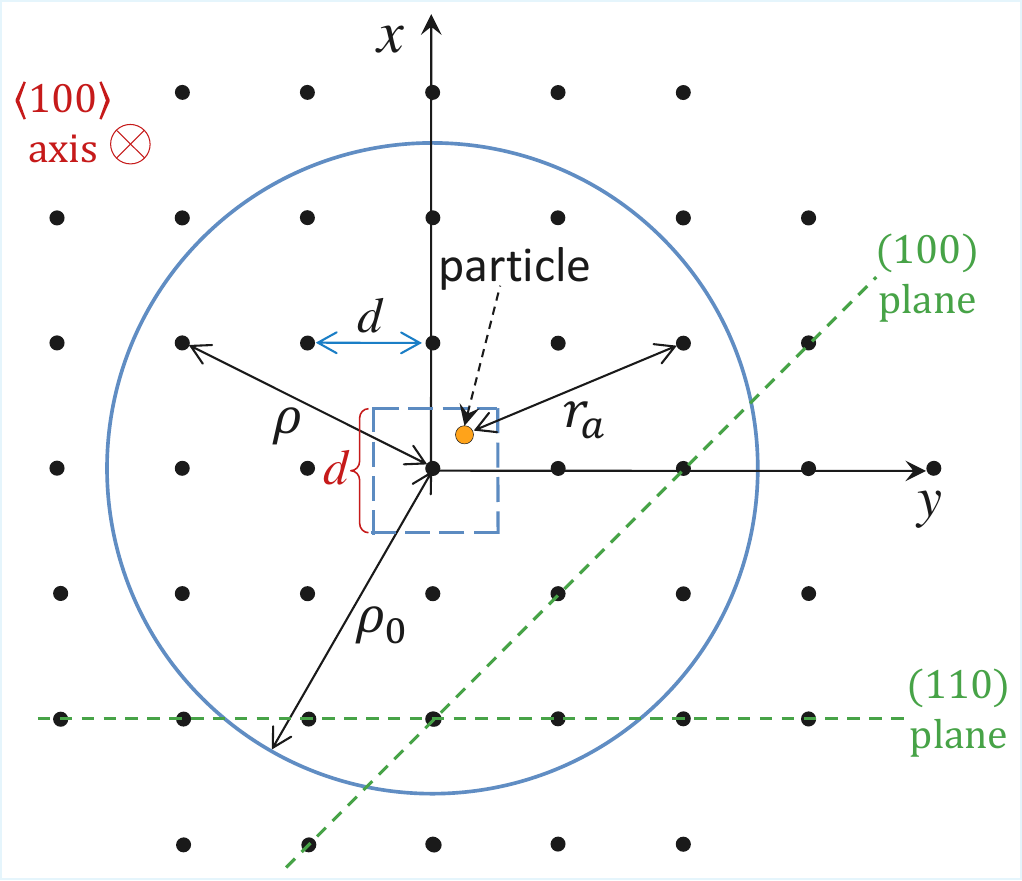}
		\caption{\label{fig1} Arrangement of atomic strings (black dots) in the plane perpendicular to $\langle100\rangle$ axis of a silicon crystal. The particle moves at a small angle with respect to $\langle100\rangle$ axis, which is parallel to the $z$ axis.}
	\end{centering}
\end{figure}

For definiteness, let us consider the case where the particle enters a silicon crystal at a small angle to the $\langle100\rangle$ axis, which is parallel to the $z$ axis. The spatial distribution of the atomic strings parallel to this axis is schematically shown in Fig.~\ref{fig1} by black dots. For simplicity, all symmetry-equivalent planes belonging to the (100) family in the silicon crystal lattice will be collectively referred to as (100) planes. In particular, this includes the (010) plane, which is denoted in Fig.~\ref{fig1} as the (100) plane. The same convention applies to the planes of the (110) family. Let us divide the atoms around the particle trajectory into two groups: the first lying inside the circular tube of radius $\rho_0$, and the second outside it. The particle interactions with the atoms from the first group are significantly influenced by the structure of the crystal lattice. The probability of the particle interaction with each of these atoms is sensitive to the precise values of the particle's coordinates $(x,y)$ inside the dashed square. The side of this square equals the distance $d$ between neighboring atomic strings along the $x$ and $y$ axes, and its center coincides with the atomic string closest to the current position of the particle. The interactions of the particle with the atoms of the second group (outside the tube) are treated in the same way as in amorphous targets. For these atoms we neglect the difference between the distance $r_a$ from the atom to the particle trajectory, and the distance $\rho$ from the atom to the $z$ axis. It is possible if $\rho_0$ is at least several times larger than $d$. 

Let us choose $\rho_0$ in such a way that the tube contains $n_0$ atomic strings inside it. The surface density of these strings in the $xy$ plane is currently $\kappa=d^{-2}$. The quantity $n_0$ equals the number of atoms inside the circle of radius $\rho_0$ within the period of the crystal lattice along the particle motion (all these atoms are seen in Fig.~\ref{fig1} without intersections). The exact value of $\rho_0$ can be determined from the requirement that the same number of atoms, $n_0$, should be contained within the considered circle in the case of their random distribution in the $xy$ plane with the same density $\kappa$. In this case our treatment will be consistent with the results for non-oriented crystals. This gives $\rho_0=(n_0/\pi\kappa)^{1/2}$. We chose $n_0=21$, as shown in Fig.~\ref{fig1}. For silicon, the lattice constant is $l\approx5.43$ \AA~and $d=l/(2\sqrt{2})\approx1.92$ \AA, which gives $\rho_0\approx5$ \AA. It is noticeably smaller than the typical transverse size $c/\omega_p$ of the particle's proper electromagnetic field, screened by the medium polarization (here $\omega_p$ is the plasma frequency of the medium), which is approximately 64 \AA. Thus, in the region $\rho<\rho_0$ there is almost no modification of the electromagnetic field associated with the TR formation process (which mainly occurs at distances on the order of $c/\omega_p$ or larger from the particle's trajectory), and the field can be considered as coinciding with the particle's Coulomb field in vacuum. It allows us to apply the following expression for the probability of an atomic K-shell electron transition from the initial ground state $|i\rangle$ to the final state $|f\rangle$ under the impact of the incident particle's field (for details see \cite{TrofymenkoKyryllin2025}):    
\begin{equation}\label{w_fi}
	w_{fi}(r)=8\alpha^2\frac{|\xi_{fi}|^2}{r^2},
\end{equation}
where $r$ is the particle impact parameter relative to the atomic center, $\xi_{fi}=\langle f|\xi|i\rangle$ is the matrix element of the atomic electron coordinate for the considered transition ($\xi$ denotes $x,y$ or $z$), and $\alpha\approx1/137$ is the fine-structure constant. Here we took into account that the K shell contains two electrons. Expression (\ref{w_fi}) is derived by treating the interaction between the incident high-energy particle and an atomic electron as a perturbation of the atomic electron Hamiltonian by the electromagnetic field of the particle moving along a straight line. Therefore, it describes the transition probability associated with a distant collision. 

Generally, the final state $|f\rangle$ can belong either to a discrete or to a continuous spectrum. In the latter case, the spectrum is formally treated as quasi-discrete. The total probability $w_0(r)$ of excitation of the K shell of an atom can be obtained by summing over all final states $|f\rangle$ with the use of the following transformation:
\begin{equation}\label{sum_x}
	\sum\limits_f|\xi_{fi}|^2=\frac{\hbar^2}{2m}\sum\limits_f F_{fi}(\hbar\omega_{fi})^{-1} =\frac{\hbar^2}{2m\bar\varepsilon},
\end{equation} 
where $F_{fi}=2m\omega_{fi}|\xi_{fi}|^2/\hbar$ can be considered as a probability distribution function for the transition $|i\rangle\to|f\rangle$ due to the sum rule $\sum\limits_f F_{fi}=1$ \cite{LandauLifshitzQM}, and $\bar\varepsilon$ is some average excitation energy of the K shell which will be estimated further [see formula (\ref{eav})]. As a result, we get:
\begin{equation}\label{w_0}
	w_0(r)=4\frac{mc^2}{\bar\varepsilon}\bigg(\frac{r_0}{r}\bigg)^2,
\end{equation}
where $r_0=e^2/mc^2$ is the classical electron radius.

In an amorphous medium, the contribution to the K-shell ionization cross section from collisions with atoms within the tube of radius $\rho_0$ can be calculated as
\begin{equation}\label{sig_p0}
	\sigma_{\rho<\rho_0}=\int\limits_{r_{\textrm{min}}}^{\rho_0} w_0(r)2\pi r dr=8\pi r_0^2\frac{mc^2}{\bar\varepsilon} \ln\frac{\rho_0}{r_{\textrm{min}}}.
\end{equation}  
Following the approach of the equivalent photon method in treating K-shell ionization \cite{Sorensen}, we chose $r_{\textrm{min}}$ to be equal to the Bohr radius of the K-shell electron's orbit:
\begin{equation}\label{r_min}
r_{\textrm{min}}=(\hbar/2m\omega_K)^{1/2},
\end{equation}  
where $\hbar\omega_K$ is the minimum energy required for ionization of the considered shell. For computer simulations, it is convenient to generalize expression (\ref{w_0}) so that it becomes valid for arbitrary $r$, as
\begin{equation}\label{w_r}
	w(r)=w_0(r)(1-e^{-r^2/b^2}).
\end{equation} 
This model for $w(r)$ reflects the fact that, at distances $r\lesssim r_{\textrm{min}}$ from the atomic center, the spatial extent of the K-shell cannot be neglected, which results in the saturation of $w(r)$ at a constant value as $r\to0$, demonstrated in \cite{Amundsen1981}. The parameter $b$ can be determined from the requirement that $w(r)$ yields the same value (\ref{sig_p0}) of the cross section $\sigma_{\rho<\rho_0}$ as $w_0(r)$. This gives:
\begin{equation}\label{b}
	b=\eta r_{\textrm{min}}e^{\Gamma/2},
\end{equation}      
where $\Gamma\approx0.5572$ is Euler's constant, and the additional factor $\eta$, which is close to unity, will be discussed below. 

Let us proceed to the case of an oriented crystal and consider the particle interaction with the atoms located inside the circle of radius $\rho_0$ contained within one period of the crystal lattice along the particle velocity (in the presently considered case of particle motion at a small angle to the $\langle100\rangle$ axis, this period coincides with the lattice constant $l$). These atoms are shown in Fig.~\ref{fig1} without intersections. The average number of the K-shell ionization events due to distant collisions of the particle with these atoms is
\begin{equation}\label{Nh}
	N_{\rho<\rho_0}=\sum\limits_{a=1}^{n_0} w(r_a),
\end{equation}  
since this number is much less than 1. Here $w(r_a)$ is defined by formula (\ref{w_r}) and the summing is performed over all the considered atoms with $\rho<\rho_0$. By analogy with an amorphous target, we can relate $N_{\rho<\rho_0}$ to the cross section for distant collisions with the considered atoms as $N_{\rho<\rho_0}=nl\sigma_{\rho<\rho_0}$, where $n$ is the number of atoms in a unit volume of the crystal. Here, $\sigma_{\rho<\rho_0}$ is the analogue of the quantity (\ref{sig_p0}) for an oriented crystal. Taking into account that the surface density of the atoms in Fig.~\ref{fig1} is $\kappa=nl$, we obtain:
\begin{equation}\label{sig_p0_crystal}
	\sigma_{\rho<\rho_0}=\frac{4r_0^2}{\kappa}\frac{mc^2}{\bar\varepsilon}\sum\limits_{a=1}^{n_0} \frac{1-e^{-r_a^2/b^2}}{r_a^2}.
\end{equation} 
The factor $\eta$ in (\ref{b}) can be defined from the requirement that, for a non-oriented crystal, the quantity (\ref{sig_p0_crystal}) should be equal to (\ref{sig_p0}). The transition to the case of a non-oriented crystal is accomplished by averaging expression (\ref{sig_p0_crystal}) over the particle coordinates $(x,y)$ within the dashed square in Fig.~\ref{fig1}. Since the number of atoms within the circle of radius $\rho_0$ is finite, the values of $r_a$ do not homogeneously cover the region $(0,\rho_0)$ during such averaging, and the factor $\eta$ may slightly differ from 1. In the considered case, the calculation gives a value of 0.995 for this factor.       

\subsection{Contribution from the region with $\rho>\rho_0$}
\label{subsec_p>p0}

Let us now consider interactions of the incident particle with the atoms outside the tube of radius $\rho_0$, which will be treated using the equivalent photon method. As noted, in this case we neglect any deviation of the particle trajectory from the $z$ axis. In this region of $\rho$, it is necessary to take into account the transformation of the electromagnetic field around the particle due to the formation of TR as the particle enters the target. The spectral density of the equivalent photon number in the field around the particle outside the tube of radius $\rho_0$ is defined as 
\begin{equation}\label{dNdw_general}
	\frac{dN_{eq}}{d\omega}=\frac{c}{4\pi^2\hbar\omega}\int\limits_{\rho_0}^{\infty}|{\bf E}_\omega|^2 2\pi\rho d\rho,
\end{equation}
where ${\bf E}_\omega$ is the Fourier component of the electric field. The cross section for K-shell ionization produced by these photons can be calculated as
\begin{equation}\label{sig_larger_p0}
	\sigma_{\rho>\rho_0}=\int\limits_{\omega_K}^{\infty}\frac{dN_{eq}}{d\omega}\sigma^K_{ph}(\omega)d\omega,
\end{equation} 
where $\sigma^K_{ph}(\omega)$ is the cross section of K-shell photoionization. The quantity ${\bf E}_\omega$ can be obtained by solving Maxwell's equations with the appropriate boundary conditions on the target surface. For normal incidence of an ultrarelativistic particle on this surface, coinciding with the $xy$ plane, in the x-ray frequency range it is given by \cite{TrofymenkoPhysRev2018}
\begin{eqnarray}\label{E_w}
	\begin{aligned}
		{\bf E}_\omega({\boldsymbol\rho}&,z)=-\frac{ie}{\pi v}\int d^2q\frac{{\bf q}e^{i{\bf q}{\boldsymbol\rho}}}{q^2+\omega^2/\gamma^2+\omega_p^2}\\
		\times&\bigg[e^{i\omega z/v}+\frac{\omega_p^2}{q^2+\omega^2/\gamma^2}e^{iz\sqrt{\omega^2-\omega_p^2-q^2}-\mu z/2}\bigg],
	\end{aligned}
\end{eqnarray}
where, for compactness, we set $c=1$ [up to the formula (\ref{dNdw_complete_density_eff})]. To restore the units, one must replace $\omega^2$ and $\omega_p^2$ with $\omega^2/c^2$ and $\omega_p^2/c^2$ in the first line and in the second term of the second line. In (\ref{E_w}), $\mu(\omega)$ is the x-ray attenuation coefficient of the target. The first term in this expression represents the proper Coulomb field of the particle, partially screened by the polarization of the medium, whereas the second term corresponds to the TR field. Substituting (\ref{E_w}) into (\ref{dNdw_general}), the squared modulus of the field can be expressed as ${\bf E}_\omega{\bf E}^*_\omega$, with each of these terms represented in the form of the integral (\ref{E_w}). After that, it is convenient to replace the lower integration limit $\rho_0$ in the integral with respect to $\rho$ in (\ref{dNdw_general}) with the upper limit $q_0=1/\rho_0$ in the integrals with respect to $q$. This corresponds to replacing the restriction in coordinate space by the corresponding restriction in momentum space (cf. the analogous procedure in \cite{TrofymenkoPhysRev2020}), since physically the quantity $\hbar q$ can be interpreted as the momentum transferred in the collision. As a result, we get:
\begin{eqnarray}\label{dNdw_precise}
	\begin{aligned}
		\frac{dN_{eq}}{d\omega}&=\frac{\alpha}{\pi\omega}\Bigg\{\Bigg[\bigg(1+\frac{2\Omega^2}{\omega_p^2}\bigg)\ln\bigg(\frac{1+q_0^2/\Omega^2}{1+q_0^2/\Omega_p^2}\bigg)-\frac{q_0^2}{q_0^2+\Omega^2}\\
		&-\frac{q_0^2}{q_0^2+\Omega_p^2}\Bigg]e^{-\mu z}+\ln\bigg(1+\frac{q_0^2}{\Omega_p^2}\bigg)-\frac{q_0^2}{q_0^2+\Omega_p^2}\\
		&+2\Lambda(\omega,q_0,z)e^{-\mu z/2}\Bigg\},
	\end{aligned}
\end{eqnarray}
where the function $\Lambda(\omega,q_0,z)$ is given by
\begin{eqnarray}\label{Lambda}
	\begin{aligned}
		&\Lambda=\frac{\Omega^2}{\omega_p^2}\bigg\{\cos\lambda_p\big[\textrm{Ci}\lambda_\gamma-\textrm{Ci}(\lambda_\gamma+\lambda_q)\big]-\textrm{Ci}(\lambda_\gamma+\lambda_p)\\
		&-\sin\lambda_p\big[\textrm{Si}\lambda_\gamma-\textrm{Si}(\lambda_\gamma+\lambda_q)\big]+\textrm{Ci}(\lambda_\gamma+\lambda_p+\lambda_q)\bigg\}\\
		&-(\lambda_\gamma+\lambda_p)\big[\textrm{Si}(\lambda_\gamma+\lambda_p+\lambda_q)-\textrm{Si}(\lambda_\gamma+\lambda_p)\big]\\
		&-\frac{\Omega_p^2}{q_0^2+\Omega_p^2}\cos(\lambda_\gamma+\lambda_p+\lambda_q)+\cos(\lambda_\gamma+\lambda_p)
	\end{aligned}
\end{eqnarray}
and $\Omega=\omega/\gamma$, $\Omega_p^2=\omega^2/\gamma^2+\omega_p^2$, $\lambda_\gamma=\omega z/(2\gamma^2)$, $\lambda_p=\omega_p^2 z/(2\omega)$, $\lambda_q=q_0^2 z/(2\omega)$. Here $\textrm{Si}(x)$ and $\textrm{Ci}(x)$ denote the sine and cosine integral functions, respectively. These functions are defined in such a way that, for $x\to+\infty$, $\textrm{Si}(x)\to\pi/2$ and $\textrm{Ci}(x)\to0$. For the value of $\rho_0$ considered here (5 \AA), one obtains $q_0=2\cdot10^7$ cm$^{-1}$. We will be interested in K-shell ionization by electrons and positrons with energies $E$ of about 1 GeV and higher. The frequencies $\omega$ that make the main contribution to (\ref{sig_larger_p0}) are on the order of $\omega_K$, due to the rapid decrease of $\sigma^K_{ph}(\omega)$ with increasing $\omega$. For silicon, $\omega_K\approx1.84$ keV, and we will restrict the integration range in (\ref{sig_larger_p0}) to $\omega_{\textrm{max}}=15$ keV. For $E=1$ GeV we have $\omega_{\textrm{max}}/\gamma\approx3.8\cdot10^5$ cm$^{-1}$, and $\omega_p\approx1.6\cdot10^6$ cm$^{-1}$. Thus, in the present case (as well as at higher particle energies), the condition $q_0\gg\omega/\gamma,~\omega_p$ holds, and expression (\ref{dNdw_precise}) can be simplified as    
\begin{eqnarray}\label{dNdw_q0}
	\begin{aligned}
		\frac{dN_{eq}}{d\omega}=\frac{2\alpha}{\pi\omega}&\Bigg\{\Bigg[\bigg(1+\frac{2\Omega^2}{\omega_p^2}\bigg)\ln\frac{\Omega_p}{\Omega}-1\Bigg]e^{-\mu z}\\
		&+\ln\frac{q_0}{\Omega_p}-\frac{1}{2}+\Lambda'(\omega,z)e^{-\mu z/2}\Bigg\},
	\end{aligned}
\end{eqnarray} 
where $\Lambda'(\omega,z)$ is defined by expression (\ref{Lambda}) without the second-to-last term containing $q_0$. Moreover, in the considered case, $(\omega_{\textrm{max}}/\gamma\omega_p)^2\ll1$, which allows the expression for $dN_{eq}/d\omega$ to be written in a very compact form:
\begin{eqnarray}\label{dNdw_w_wpy}
	\begin{aligned}
		&\frac{dN_{eq}}{d\omega}=\frac{2\alpha}{\pi\omega}\Bigg\{\ln\frac{q_0}{\omega_p}-\frac{1}{2}+\bigg[\ln\frac{\gamma\omega_p}{\omega}-1\bigg]e^{-\mu z}\\
		&+\bigg[\cos(\lambda_\gamma+\lambda_p)-\lambda_p\big[\pi/2-\textrm{Si}(\lambda_\gamma+\lambda_p)\big]\bigg]e^{-\mu z/2}\Bigg\}.
	\end{aligned}
\end{eqnarray} 
For $z\to0$, this expression takes the form:
\begin{equation}\label{dNdw_no_density_eff}
	\frac{dN_{eq}}{d\omega}=\frac{2\alpha}{\pi\omega}\Bigg\{\ln\frac{q_0\gamma}{\omega}-\frac{1}{2}\Bigg\}
\end{equation} 
and coincides with the equivalent photon spectrum of a high-energy particle in vacuum. In this case, the density effect is entirely absent in K-shell ionization. For $z\gg\mu^{-1}(\omega_\textrm{max})$, the TR photons with frequencies contributing to the cross section (\ref{sig_larger_p0}) are already absorbed in the medium, and the spectrum is given by
\begin{equation}\label{dNdw_complete_density_eff}
	\frac{dN_{eq}}{d\omega}=\frac{2\alpha}{\pi\omega}\Bigg\{\ln\frac{q_0}{\omega_p}-\frac{1}{2}\Bigg\}.
\end{equation}
In this case, the density effect reaches its asymptotic magnitude and the cross section (\ref{sig_larger_p0}) does not depend on the particle energy.

When the crystal is rotated, the condition of normal particle incidence on the crystal surface is no longer satisfied. However, the range of rotation angles, which we will be interested in, is on the order of 1 mrad for a particle energy of 1 GeV and becomes even smaller at higher energies. Moreover, at high particle energies, the properties of x-ray TR in the forward direction are rather insensitive to the angle between the particle velocity and the target surface, even for much larger deviations of this angle from $90^\circ$ \cite{Garibyan1960}. Thus, the derived formulae for $dN_{eq}/d\omega$ are valid within the considered ranges of crystal rotation. In particular, within these ranges, the $z$-axis, which rotates together with the crystal, remains almost parallel to the initial direction of the incident particle motion.

The total cross section of K-shell ionization in an oriented crystal due to distant collisions is the sum of the contributions (\ref{sig_p0_crystal}) and (\ref{sig_larger_p0}): 
\begin{equation}\label{sig_d}
	\sigma_d=\sigma_{\rho<\rho_0}+\sigma_{\rho>\rho_0}.
\end{equation} 
In the first term of this expression, the parameter $\overline{\varepsilon}$ can be determined by comparing the quantity (\ref{sig_p0}) in a non-oriented crystal with the corresponding result obtained using the equivalent photon method:
\begin{equation}\label{sig_p0_eqv_ph}
	 \sigma_{\rho<\rho_0}=\int\limits_{\omega_K}^{\infty}\frac{dN'_{eq}}{d\omega}\sigma^K_{ph}(\omega)d\omega,
\end{equation} 
where $dN'_{eq}/\omega$ is the equivalent photon spectral density associated with the distances $r_{\textrm{min}}<r<\rho_0$ from the particle trajectory. It can be calculated as $dN_{eq}^t/d\omega-dN_{eq}/d\omega$, where $dN_{eq}^t/d\omega$ is the total spectral density of equivalent photons at $r>r_{\textrm{min}}$ , whereas $dN_{eq}/d\omega$ is the spectral density in the expression (\ref{sig_larger_p0}). The quantity $dN_{eq}^t/\omega$ can be obtained from (\ref{dNdw_w_wpy}) by the substitution of $q_0=1/\rho_0$ with $q_1=1/r_\textrm{min}$. Since $q_1\gg q_0$, the applicability conditions of (\ref{dNdw_w_wpy}) are not violated in this case. Thus, $dN'_{eq}/\omega$ is given by
\begin{equation}\label{dN'dw}
		\frac{dN'_{eq}}{d\omega}=\frac{2\alpha}{\pi\omega}\ln\frac{\rho_0}{r_{\textrm{min}}}=\frac{2\alpha}{\pi\omega}\ln\bigg(\rho_0\sqrt{\frac{2m\omega_K}{\hbar}}\bigg).
\end{equation} 
This quantity does not depend on $z$, since we neglect any field transformation in the region $\rho<\rho_0$. Substituting (\ref{dN'dw}) to (\ref{sig_p0_eqv_ph}) and comparing this expression with (\ref{sig_p0}), we get
\begin{equation}\label{eav}
	\overline{\varepsilon}=\frac{4\pi^2 e^2\hbar}{mc} \bigg(\int\limits_{\omega_K}^{\infty}\sigma^K_{ph}(\omega)\frac{d\omega}{\omega}\bigg)^{-1}.
\end{equation} 
For the values of $\omega$ contributing to the integrals in (\ref{sig_larger_p0}) and (\ref{eav}), photon attenuation occurs predominantly due to atomic photoionization. This allows the numerical values of $\sigma_{ph}^K$ to be determined based on the photon attenuation lengths $\mu^{-1}$ \cite{XrayWebSite}, using the relation $\sigma_{ph}^K=p_K\mu/n$. Here $p_K=(J_K-1)/J_K$, where $J_K\approx125/Z+3.5$ is the scaling factor \cite{Sorensen} that determines the contribution of K-shell electrons to the total photoionization cross section of the atom, and $Z$ denotes the atomic number of the medium. From (\ref{eav}), we obtain the numerical value $\overline{\varepsilon}\approx3915$ eV.

On the basis of (\ref{sig_p0_crystal}), (\ref{sig_larger_p0}), (\ref{sig_d}) and (\ref{eav}) it is possible to calculate the number $N_d=\sigma_dn$ of K-shell ionization events per unit path length due to distant collisions in an oriented crystal: 
\begin{eqnarray}\label{N_d}
	\begin{aligned}
		N_d&=4\pi nr_0^2\frac{mc^2 }{\overline{\varepsilon}}\Bigg\{\frac{1}{\pi\kappa}\sum\limits_{a=1}^{n_0} \frac{1-e^{-r_a^2/b^2}}{r_a^2}\\
		&+\frac{\pi}{\alpha}\bigg(\int\limits_{\omega_K}^{\infty}\sigma^K_{ph}(\omega)\frac{d\omega}{\omega}\bigg)^{-1}\int\limits_{\omega_K}^{\infty}\frac{dN_{eq}}{d\omega}\sigma^K_{ph}(\omega)d\omega\Bigg\}.
	\end{aligned}
\end{eqnarray} 
\begin{figure}
	\begin{centering}
		\includegraphics[width=0.5\textwidth]{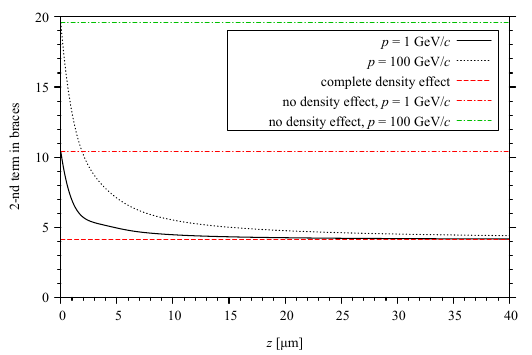}
		\caption{\label{fig2}Dependence of the second term in the braces of (\ref{N_d}) on the distance from the surface of a silicon target for two electron (positron) momentum values.}
	\end{centering}
\end{figure} 
Here, $N_d$ is written in a convenient form, where the expression in braces is dimensionless. The quantity $dN_{eq}/d\omega$ is given by expression (\ref{dNdw_w_wpy}). The first term in (\ref{N_d}) depends on the particle's position relative to the $n_0$ atoms closest to its trajectory and, hence, is a function of the crystal orientation. The second term does not depend on this orientation. For a fixed $\rho_0$, it is a function of the particle energy $E$ and of its path length $z$ inside the crystal. Fig.~\ref{fig2} shows the dependence of this term on $z$ for two values of the incident electron or positron momenta. The value of this term for $z\to0$ corresponds to $dN_{eq}/d\omega$ defined by the expression (\ref{dNdw_no_density_eff}) and increases with increasing particle momentum. The asymptote at large $z$, on the contrary, is determined by the spectrum (\ref{dNdw_complete_density_eff}) and is the same for different momenta of high-energy electrons and positrons. The value $\mu^{-1}(\omega_\textrm{max})$ is approximately 450 $\mu$m. The figure shows that the second term closely approaches its asymptotic value at much smaller $z$. It indicates that the value of $\omega_\textrm{max}$ was chosen with a good margin, and the main contribution to (\ref{N_d}) comes from significantly smaller $\omega$. For comparison, the attenuation length $\mu^{-1}$ in the immediate vicinity of $\omega_K$ (for $\omega>\omega_K$) is around 1.3 $\mu$m. Fig.~\ref{fig2} shows that the second term in (\ref{N_d}) undergoes a significant change due to the process of TR formation and attenuation in a region extending up to 10 $\mu$m or more from the surface, where $N_d$ differs from its values both with and without the density effect. This region is comparable to the thickness of the boundary layer from which CXR can escape the target (for the silicon $K_\alpha$ line, $\mu^{-1}$ is approximately 12 $\mu$m). This highlights the importance of accounting for the considered transformation of the electromagnetic field within the target's boundary layer for high-energy particles.

\subsection{Close collisions}
\label{subsec_Close_collisions}

Let us now consider the contribution to K-shell ionization arising from close collisions. It can be derived based on the model of binary collisions between the incident particle and atomic electrons. For incident electrons, this can be done using the M{\o}ller cross section. For ultrarelativistic electrons, in the region of energy transfers $\varepsilon$ that make the main contribution to the total collision cross section, it takes the form of the Rutherford cross section:
\begin{equation}\label{dif_sig_close}
	\frac{d\sigma_c}{d\varepsilon}=\frac{2\pi e^4}{mc^2 \varepsilon^2}
\end{equation}   
and can be applied to both electrons and positrons. Following the approaches adopted in \cite{Sorensen,Bak}, we apply this form of $d\sigma_c/d\varepsilon$ down to the ionization threshold of the K shell. In this case, the total cross section for ionization of the K shell, which contains two electrons, due to close collisions is given by
\begin{equation}\label{sig_close}
	\sigma_c=\int\limits_{\hbar\omega_K}^{\varepsilon_{\textrm{max}}}\frac{d\sigma_c}{d\varepsilon}d\varepsilon=\frac{4\pi e^4}{mc^2\hbar\omega_K},
\end{equation}   
since the maximum energy transfer $\varepsilon_{\textrm{max}}$ significantly exceeds $\hbar\omega_K$. For silicon, the numerical value of this cross section is 277 barns, which is much smaller than the typical values of $\sigma_d$ in a non-oriented crystal [at $E=1$ GeV, $\sigma_c$ is roughly 11\% of the $\sigma_d$ value (\ref{sig_d}) in the absence of the density effect]. It is also much smaller than the classical area of the K shell, which is roughly on the order of $10^6$ barns. Therefore, as in the simulations of ionization energy loss in \cite{TrofymenkoKyryllin2024,TrofymenkoKyryllin2025}, we assume that the number $N_c$ of K-shell ionization events per unit path length due to close collisions in an oriented crystal is determined by the local concentration $n({\boldsymbol\rho})$ of the K-shell electrons divided by 2 and averaged over the $z$ coordinate:
\begin{equation}\label{N_c}
	N_c=4\pi n({\boldsymbol\rho})r_0^2\frac{mc^2 }{\hbar\omega_K}.
\end{equation}

\subsection{Simulation of CXR photon number}
\label{subsec_CXR}

By simulating particle trajectories in an oriented crystal and using expressions (\ref{N_d}) and (\ref{N_c}), one can determine the total number of K-shell ionization events per unit length of the particle trajectory. The angular density of the number of CXR photons emitted from the upstream surface of the target is then given by
\begin{equation}\label{dNdo}
	\frac{dN}{do}=\frac{w_fP_\alpha}{4\pi}\int(N_d+N_c) e^{-\mu(\omega_\textrm{CXR})z/\cos\vartheta}dz.
\end{equation}
Here, $\vartheta$ is the angle between the observation direction and the direction opposite to the initial direction of the incident particle motion, and $w_f=0.05$ is the K-shell fluorescence yield for silicon. We will consider the $K_\alpha$ line, which dominates the CXR spectrum of silicon. The quantity $P_\alpha$ in (\ref{dNdo}) represents the relative probability of $K_\alpha$ line emission during the radiative transition that fills a vacancy in the K-shell. In the considered case $P_\alpha\approx0.987$. We will consider the entire $K_\alpha$ line without distinguishing between the $K_{\alpha 1}$ and $K_{\alpha 2}$ components. The CXR frequency is taken to be $\omega_\textrm{CXR}=1739.38$ eV, corresponding to that of the $K_{\alpha 1}$ line.

\section{Simulation results}
\label{sec3} 

In the present section, using the simulation method for characteristic X-ray radiation produced by high-energy charged particles passing through a crystal, presented above, we investigate the dependence of $dN/do$ on the crystal orientation with respect to the direction of motion of the incident particles.
In our analysis, a silicon crystal was chosen as the target. This choice is motivated by the fact that silicon crystals are most commonly used in experimental studies of effects related to particle channeling in oriented crystals.
The reason is that, on the one hand, silicon has a diamond-type lattice, and the dense packing of atoms in this lattice results in strong intracrystalline fields of atomic strings and planes. On the other hand, silicon-crystal growth technology is now highly advanced, enabling the production of crystals with extremely low defect, dislocation, and impurity levels.

The simulation was based on the numerical solution of the equations of motion for relativistic electrons and positrons in a crystal oriented along atomic planes or atomic strings.
The electric potential of these planes and strings was calculated using the continuous potential approximation \cite{Lindhardt1965}.
For the atomic potential, the Doyle-Turner approximation \cite{doyle_relativistic_1968} was used.
In this approximation, potential energy of an electron in the field of the (100) atomic planes of a silicon crystal can be written as \cite{kyryllin_influence_2018}:
\begin{equation}\label{U_p}
	U_{pl}(\xi) = -\frac{2 \pi \hbar^2}{m d_s d_a d_p} \sum_{k=1}^{4} \alpha_k 
	\theta_3	\left(
		\pi \frac {\xi} {d_p} ,
		\exp\left(-\frac {\beta_k + B} {4 d_p^2}\right)
	\right),
\end{equation}
where $\xi$ is the distance from the plane, $m$ is an electron mass, $d_p$ is the distance between neighboring atomic planes (100), $d_s$ is the distance between neighboring atomic strings in the atomic plane, $d_a$ is the distance between neighboring atoms in these strings, $\alpha_k$ and $\beta_k$ are coefficients found in \cite{doyle_relativistic_1968} for a large number of elements, $B = 8 \pi^{2} r_{_{T}}^{2}$ and $r_{_{T}}$ is the rms atomic thermal vibration amplitude in one direction ($r_{_{T}} \approx 0.075$~{\AA} for Si at 293\thinspace K), $\theta_3 (u , q) = \sum_{n = -\infty}^\infty \limits q^{n^2} \cos(2 n u)$ is the Jacobi theta function of the third kind \cite{Gradshteyn}.

If an electron moves in the continuous potential of the atomic strings of the silicon crystal along the $\langle 100 \rangle$ direction, its potential energy can be found as
\begin{eqnarray}\label{U_s}
	\begin{aligned}
		U_{ax} \left(x,y\right) =
		&-\frac {2 \pi \hbar^2} {m d_a d^2} \sum_{k=1}^4 \limits \alpha_k
		\theta_3	\left(
			\pi \frac x {d} ,
			e^{-\frac {\beta_k + B} {4 d^2}}
		\right)\\
		\times&\theta_3 \left( 
			\pi \frac y {d} ,
			e^{-\frac {\beta_k + B} {4 d^2}}
		\right),
	\end{aligned}
\end{eqnarray}
where $x$, $y$ and $d$ are shown in Fig.~\ref{fig1}. In the simulation of charged particle trajectories, scattering on thermal vibrations of crystal atoms and on atomic electrons was taken into account, as described in \cite{shul2013stochastic}.

\begin{figure}
	\begin{centering}
		\includegraphics[width=0.5\textwidth]{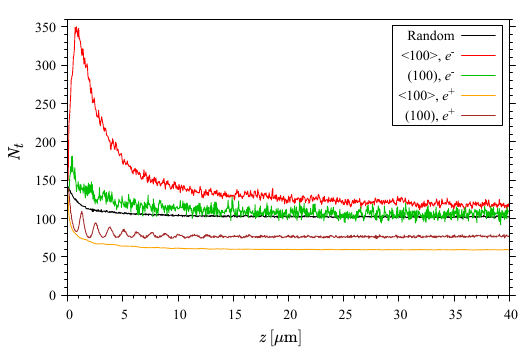}
		\caption{\label{fig3} Dependence of $N_t=N_d + N_c$ on the particle path length in the crystal for different crystal orientations.}
	\end{centering}
\end{figure}

Let us begin the analysis of the simulation results with the case of electron and positron energies of 1 GeV.
Fig.~\ref{fig3} shows the dependence of the total number of K-shell ionization events per unit path $N_t=N_d + N_c$ on the path length $z$ of the incident particle inside the crystal for different crystal orientations.   
The number of particles used in the simulation was $10^5$.
The black line corresponds to the case of a ``random" crystal orientation, i.e., when the direction of particle motion makes a large angle with the crystal axes and planes, and the particle undergoes multiple scattering similarly to that in an amorphous target (hereinafter, we shall use the terms ``amorphous target" or ``random crystal orientation" for this case).
The behavior of this curve is determined by the dependence of the second term in (\ref{N_d}) on $z$, shown in Fig.~\ref{fig2}. Currently, the difference between the values of $N_t$ at $z=0$ and at large $z$ is less pronounced than in Fig.~\ref{fig2} due to the contributions from the first term in (\ref{N_d}) and from $N_c$, but it increases with the particle energy. 

The green curve corresponds to the motion of electrons in a silicon crystal oriented along the (100) planes with respect to the incident particle momentum.
The enhancement of $N_t$ compared to the random crystal orientation is caused by planar channeling of electrons, where the probability of collisions with K-shell electrons increases due to the attractive force between the incident electrons and the periodically arranged atomic nuclei.
For the same reason, the dechanneling length of negatively charged particles is significantly shorter than that of positively charged ones, since scattering on thermal vibrations of atoms, which leads to dechanneling, becomes more intense as the probability of particle-atom collisions increases.
Rapid dechanneling causes the green curve to quickly approach the black one, corresponding to the transition of initially channeled particles to the over-barrier mode of motion as they move deeper inside the crystal.

The brown curve corresponds to the incidence of positrons on the crystal oriented along the (100) planes.
Due to the repulsive force acting between the positrons and atomic nuclei, collisions with K-shell electrons are suppressed during the motion of these particles in the field of atomic planes, resulting in $N_t$ being lower than in the amorphous target.
The oscillations at small crystal thicknesses are related to the fact that the potential of atomic planes deviates significantly from the parabolic shape only at small distances from the planes.
Most of the channeled positrons do not reach this region while moving in the crystal. Therefore, their oscillation frequencies in the (100) planar channel are nearly identical, since in a parabolic potential, the oscillation frequency is independent of the impact parameter.
The differences in the oscillation frequencies of positrons in the planar channel become more pronounced with increasing the particle path length in the crystal, resulting in the smearing of the oscillations in Fig.~\ref{fig3} at large $z$.

The yellow curve corresponds to the incidence of positrons on the crystal with the axial orientation $\langle 100 \rangle$.
In the axial channeling mode, collisions of positrons with atomic electrons are also suppressed, so again we observe a decrease in $N_t$ compared to the amorphous target.
Finally, the red curve corresponds to axial channeling of electrons, which leads to a significant increase in $N_t$ at small $z$. The maximum of this curve is slightly shifted from the crystal surface and corresponds to the value of $z$ at which the average distance between the incident electrons and the atomic strings, nearest to them, reaches its minimum.
Due to intense dechanneling, $N_t$ rapidly decreases with increasing $z$, approaching the value corresponding to the amorphous target. However, this value is not reached by $N_t$ in the considered range of $z$. Since incoherent scattering can not only increase the transverse energy of particles in a crystal but also decrease it, it can induce both transitions of electrons from under-barrier states (channeling) to above-barrier states, referred to as dechanneling, and transitions from above-barrier states to under-barrier ones, referred to as rechanneling. It is precisely rechanneling that prevents the value of $N_t$ at a crystal thickness of 40~$\mu$m for axial orientation from reaching the value characteristic of a randomly oriented crystal of the same thickness. A similar influence of rechanneling can be observed for planar channeled electrons, as the green curve in Fig.~\ref{fig3} remains slightly higher than the black one within the considered range of $z$. However, in this case the effect is much weaker. It should be noted that at $z=0$, all the curves converge to the value characteristic of the amorphous target. This is because the coordinates of the particle entrance into the crystal are randomly distributed in the $xy$ plane, making the ionization process at $z\to0$ analogous to that in the amorphous target.

\begin{figure}
	\begin{centering}
		\includegraphics[width=0.5\textwidth]{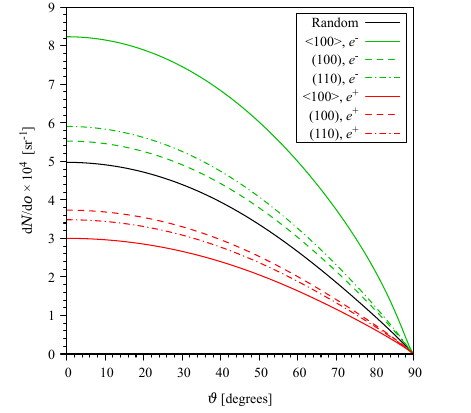}
		\caption{\label{fig4} Dependence of angular density of the number of CXR photons emitted from the upstream surface of the crystal on the angle between the observation direction and the direction opposite to the $z$ axis.}
	\end{centering}
\end{figure}

Fig.~\ref{fig4} shows the angular density of the number of CXR photons emitted from the upstream surface of the crystal as a function of the angle $\vartheta$ between the observation direction and the direction opposite to the $z$ axis for different orientations of the crystal, whose thickness was chosen to be 40 microns. The calculation is based on the formula (\ref{dNdo}).
As expected from the results shown in Fig.~\ref{fig3}, positrons traversing an oriented crystal exhibit a decrease in $dN/do$ relative to its value in the amorphous target. This decrease is caused by the suppression of the probability of both close and distant collisions of the particles with K-shell electrons in this case.
For electrons, on the contrary, the increased probability of such collisions in the channeling mode, leads to an increase in $dN/do$ compared to the case of particle motion in the amorphous target.
With increasing $\vartheta$, photons travel a longer path in the crystal [see the exponent in (\ref{dNdo})], which results in stronger attenuation of CXR and its total extinction as $\vartheta\to90^\circ$.

In addition to the planar orientation (100), Fig.~\ref{fig4} also shows the case of the planar orientation (110).
The potential well in which a positron moves during planar channeling is about 11.8 eV deep for the (100) orientation, whereas for the (110) orientation it is approximately 21.4 eV.
Moreover, the (110) planar channel is $\sqrt{2}$ times wider than the (100) channel (see Fig.~\ref{fig1}).
The wider and deeper potential well leads to a stronger suppression of the collision probability for positrons in the (110) orientation compared to that in the (100) orientation.
For electrons, on the contrary, the probability of collisions with K-shell electrons is higher for the (110) orientation than for (100).
This explains why, in Fig.~\ref{fig4}, the curves corresponding to the (110) orientation lie farther from the black line, which represents the case of an amorphous target, than those corresponding to the (100) orientation.

\begin{figure}
	\begin{centering}
		\includegraphics[width=0.5\textwidth]{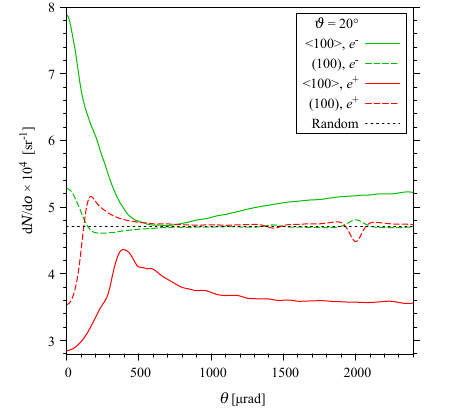}
		\includegraphics[width=0.5\textwidth]{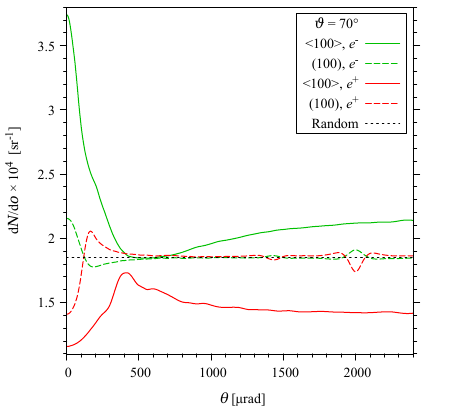}
		\caption{\label{fig5} Dependence of the angular density of the number of CXR photons emitted from the upstream surface of the crystal on the angle between the atomic axis/plane and the momentum of the particles incident on the crystal. The particle energy is 1 GeV.}
	\end{centering}
\end{figure}

Now let us fix the observation angle $\vartheta$ and consider the dependence of the angular density of the number of CXR photons on the crystal orientation. 
Fig.~\ref{fig5} (top) shows the dependence of $dN/do$ on the angle $\theta$ at $\vartheta = 20^\circ$.
For the planar orientation (100), $\theta$ is the angle between the (100) plane and the momentum of the particles incident on the crystal.
For the axial orientation $\langle 100 \rangle$, $\theta$ is the angle between the $\langle 100 \rangle$ axis of the crystal and the particle momentum. In the latter case, the direction of crystal rotation is chosen such that, with increasing $\theta$, the crystal gradually approaches the (100) planar orientation [the rotation axis is perpendicular to the (100) plane in Fig.~\ref{fig1}].
In this case, the transition from the axial to the planar orientation occurs at angles $\theta$ much larger than the critical angle for axial channeling, which, for the considered particle energy of 1 GeV, is about 385 $\mu$rad.

It can be seen that, for both positrons and electrons, the transition from the planar channeling regime to over-barrier motion corresponds to a non-monotonic change in $dN/do$. 
For positrons, at an angle $\theta$ equal to the critical planar channeling angle (which is roughly half the critical axial channeling angle), $dN/do$ reaches a maximum, in which the photon yield exceeds the yield $(dN/do)_R$ characteristic of the randomly oriented target. For electrons, the photon yield reaches a minimum at this angle, where it is slightly smaller than $(dN/do)_R$.
These extrema are associated with the so-called ``hanging-over" effect in the particle motion in the crystal. The manifestation of this effect has also been investigated for the most probable ionization energy loss of particles in crystals \cite{TrofymenkoKyryllin2024}. It is associated with the fact that, when a positively charged particle enters a crystal at an angle close to the critical angle of planar channeling, it slows down in the transverse direction near an atomic plane and spends a significant amount of time in the region of enhanced atomic electron density, which increases the probability of the particle collisions with the electrons. A negatively charged particle, on the contrary, slows down in the middle between neighboring planes (the region of decreased atomic electron density), which results in a decreased probability of collisions. 
It should also be noted that the small extrema near $\theta=2$~mrad correspond to planar channeling in high-index planes, which in the present case is (150) plane.

In the case of the axial crystal orientation, a non-monotonic variation of $dN/do$ with increasing angle between the particle momentum and the ⟨100⟩ axis also occurs. 
However, in this case, the non-monotonicity is mainly due to the fact that when $\theta$ exceeds the critical angle for axial channeling, particles begin to be captured into the planar channeling regime in the field of atomic planes (100), which, in the limit of large angles, causes $dN/do$ to approach the value characteristic of planar channeling. In the intermediate range of $\theta$, the photon yield approaches (or even reaches) the value $(dN/do)_R$ characteristic of the amorphous target, which is higher (for positrons) or lower (for electrons) than the values typical of both axial or planar channeling.

Fig.~\ref{fig5} (bottom) shows the same dependence of $dN/do$ on $\theta$, but for an observation angle $\vartheta = 70^\circ$.
It can be seen that the qualitative behavior of the curves remains the same, while the quantitative values differ. 
This difference is due to the increased attenuation of CXR with increasing $\theta$. It causes not only a decrease in the absolute values of the photon yield, but also an increase in the magnitude of the orientational effects for electrons. Indeed, in the present case, for instance, $dN/do$ for axially channeled electrons at $\theta=0$ is twice as large as $(dN/do)_R$, whereas for $\vartheta = 20^\circ$ the ratio of these quantities is approximately 1.67. The reason for this is that at $\vartheta = 70^\circ$, the photons are effectively emitted from a thinner boundary layer of the crystal than at $\vartheta = 20^\circ$, and the dechanneling effects, which are crucial for electrons, are manifested to a lesser extent. 

To give a sense of the time scale required to perform the simulations whose results are presented in this paper, we note that the total computation time needed to generate the plots shown in Fig.~\ref{fig5} was dominated by the calculation of trajectories of $10^5$ charged particles in an oriented silicon crystal of 40 $\mu$m thickness, since the evaluation of $dN/do$ is significantly faster than the trajectory calculations.
Since the particles in the beam are assumed to be non-interacting with each other, the trajectories of individual particles can be computed independently and therefore in parallel.
As a result, the total computation time scales inversely with the number of computational cores used.
For a quantitative estimate, we note that on a single CPU core with a clock frequency of 4 GHz, the computation of the plots shown in Fig.~\ref{fig5} would require approximately 238 hours.
Accordingly, performing the same calculations on 16 CPU cores would reduce the required time to slightly less than 15 hours.   

\begin{figure}
	\begin{centering}
		\includegraphics[width=0.5\textwidth]{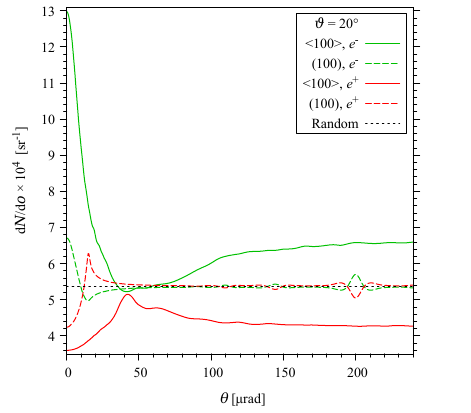}
		\caption{\label{fig6} The same as in Fig.~\ref{fig5}, but for the particle energy of 100 GeV.}
	\end{centering}
\end{figure}

Let us now analyze how the dependencies shown in Fig.~\ref{fig5} change with increasing particle energy.
Fig.~\ref{fig6} presents the dependence of $dN/do$ on $\theta$ for $\vartheta = 20^\circ$ for electron and positron energy of 100 GeV.
The critical channeling angles (both axial and planar) depend on particle energy as $E^{-1/2}$. 
Therefore, when the particle energy increases from 1 GeV to 100 GeV, these angles decrease by a factor of 10.
As Fig.~\ref{fig6} shows, the qualitative behavior of the $dN/do$ curves does not change with increasing energy; however, the angles at which the local extrema in the plotted curves occur decrease by a factor of 10. The figure also shows that the effects of crystal orientation in CXR at 100 GeV are more pronounced than at 1 GeV. For instance, the ratio of $dN/do$ to $(dN/do)_R$ at $\theta=0$ for axially channeled electrons is about 2.4 in the present case. This might seem somewhat unexpected, since the second term in (\ref{N_d}), which does not depend on the crystal structure, increases with particle energy. However, the dechanneling length also increases, enhancing the difference between the contributions from the first term in (\ref{N_d}), as well as from $N_c$, and their contributions in an amorphous target in the boundary layer of the crystal responsible for CXR emission.

\begin{figure}
	\begin{centering}
		\includegraphics[width=0.5\textwidth]{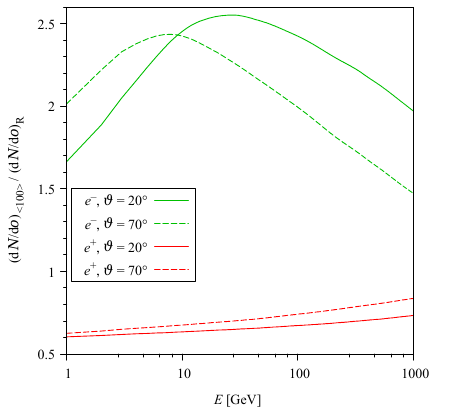}
		\caption{\label{fig7} Dependence of the angular density of the number of CXR photons emitted from the upstream surface of the crystal, normalized by its value in the amorphous target, on the particle energy.}
	\end{centering}
\end{figure}

Finally, Fig.~\ref{fig7} presents the overall evolution of the CXR yield with particle energy in an axially oriented crystal over a wide energy range. The figure depicts the ratio of the radiation angular density $dN/do$ in the crystal and the corresponding angular density $(dN/do)_R$ in the amorphous target. The results are presented both for electrons and positrons for two values of the angle $\vartheta$: $20^\circ$ and $70^\circ$.
It can be seen that for positrons, this ratio increases monotonically with energy, whereas for electrons it first increases and then decreases.
This behavior is determined by two competing factors, which were mentioned in connection with Fig.~\ref{fig6}.
On the one hand, the second term in expression (\ref{N_d}) for $N_d$ increases with energy due to the relativistic increase of the transverse scale of the particle's electromagnetic field at the crystal entrance, bringing $N_d$ closer to its value in the amorphous target. On the other hand, the dechanneling length also increases with particle energy, causing the first term in (\ref{N_d}) and the quantity $N_c$ to deviate further from their values in the amorphous targets over a larger portion of the boundary layer of the crystal that contributes to the CXR yield.
For not very high electron energies, the first of these factors dominates, leading to an increase in the relative CXR yield. As the electron energy increases further, the dechanneling length becomes larger than the thickness of the above-mentioned boundary layer, and its subsequent growth no longer affects the CXR yield. In this case, only the second factor plays a role, and the relative radiation yield decreases, tending toward unity. At $\vartheta=70^\circ$, the maximum of the curve corresponds to a lower particle energy than at $\vartheta=20^\circ$, since the effective thickness of the boundary layer is smaller in this case. 

It is also worth noting that, as the analysis shows, the maximum of the red curve in Fig.~\ref{fig2} shifts to higher $z$ values (for which the emitted CXR is more strongly attenuated) as the electron energy increases. It is the result of the fact that the electron trajectories become increasingly rectilinear, and the region where each particle approaches the atomic string shifts to larger values of $z$. On the one hand, this factor is one of the reasons for the increase in the dechanneling length with energy. On the other hand, it may play an independent role in decreasing the ratio $(dN/do)/(dN/do)_R$ for high-energy electrons, when the dechanneling length already exceeds the thickness of the boundary layer. For positrons, the influence of the dechanneling process is much weaker than for electrons, which results in a monotonic approach of their CXR yield to the value characteristic of the amorphous target as the particle energy increases.

From Figs.~\ref{fig5}-\ref{fig7}, we see that detecting CXR generated by high-energy electrons and positrons in oriented crystals, it is possible to obtain valuable information about the properties of particle motion in the crystal and the evolution of these properties with the change of the crystal orientation. In addition to the fundamental interest, this also opens up various practical applications of this phenomenon. In particular, it may include the development of noninvasive methods for finding an optimal crystal orientation relative to the incident beam direction, as proposed in earlier works, as well as the experimental investigation of the dechanneling process. Indeed, Fig.~\ref{fig7} makes it possible to roughly estimate the electron dechanneling length in the axial crystal orientation for a certain particle energy, at which this length becomes comparable to the effective thickness of the boundary layer contributing to the CXR yield. Other geometries may allow a more precise quantitative investigation of this process in a wide range of particle energies. For instance, this may involve detection of CXR from a lateral surface of the crystal, parallel to the direction of the particle beam. These possibilities are planned to be studied elsewhere. 

It is also worth noting that results resembling those presented in this paper might also be expected for a germanium crystal, which has the same lattice structure as silicon. Such crystals are also applied in the experiments with high-energy particle beams. On the one hand, they are much more expensive than silicon crystals. On the other hand, they can produce more intense CXR with a higher frequency and, hence, weaker attenuation.

Finally, note that in the present paper we neglected the angular divergence of the incident particle beam, aiming to study the `pure' manifestation of the crystal orientation effects on K-shell ionization and CXR. In a real experiment, this corresponds to using a beam with a divergence angle significantly smaller than the critical angle of planar or axial channeling, or to using a coincidence scheme that allows one to register the signal only from well-channeled particles, such as the one employed in \cite{Esbensen1978}.   

Among the facilities where the discussed orientational effects in CXR could be experimentally investigated is the MAMI microtron in Mainz. It can provide a high-quality electron beam with energies on the order 1 GeV (specifically, up to 855 MeV) and an angular divergence of about 20 $\mu$rad or less, depending on the beam spot size \cite{Backe2008}, which is much smaller than the critical angles $\psi_c$ and $\theta_c$ for axial and planar channeling for the crystal orientations considered at such particle energies. A high-quality positron beam of a slightly lower energy is also available at this facility. For the multi-GeV energy range (up to about 150 GeV for electrons and positrons), possible options include the H2 and H4 beam lines at the CERN SPS accelerator. Here, the achievable beam angular divergence is around 100 $\mu$rad and exceeds the corresponding $\psi_c$ and $\theta_c$ values. However, with the use of an appropriate tracking system, it is possible to select particles moving within angles of about several $\mu$rad, which are significantly smaller than the angles $\psi_c$ and $\theta_c$ in this case. For an application of such a system at this facility, see, e.g., \cite{Soldani2025}.

\section{Conclusions}
\label{Conclusions}

The paper presents the results of an investigation of atomic K-shell ionization and the characteristic x-ray radiation (CXR) produced by high-energy electrons and positrons in oriented silicon crystals. The investigation is carried out on the basis of computer simulations of particle trajectories inside the crystal and the process of K-shell ionization. A method for simulating this ionization process has been developed, taking into account the transformation of the particle's electromagnetic field upon its entrance into the crystal due to the formation of transition radiation, leading to the onset of the density effect in the ionization. The dependence of the number of K-shell ionization events per unit path on particle path length inside the crystal is presented for the axial $\langle100\rangle$ and planar (100) orientations of the crystal with respect to the particle beam direction. The angular density of the number of CXR photons $dN/do$, emitted from the upstream surface of the crystal, is calculated for these orientations, as well as for the (110) planar orientation. The evolution of $dN/do$ with increasing angle $\theta$ between the particle momentum and the $\langle100\rangle$ axis or the (100) plane is investigated. In the first case, the crystal rotation changing the orientation from the axial $\langle100\rangle$ to the planar (100) is considered. It is demonstrated that, both for electrons and positrons, the change in $dN/do$ during these transitions is not monotonic. For the transitions from the planar orientation, the ``hanging-over" effect in particle motion can manifest itself in K-shell ionization and CXR emission. According to the obtained results, under the considered conditions the influence of crystal orientation is most pronounced for electrons in the axial orientation of the crystal. The evolution of $dN/do$ with changing particle energy is investigated over a wide range of energies, from 1 GeV to 1 TeV, for the axial crystal orientation $\langle100\rangle$. It is shown that for positrons, with increasing particle energy, the radiation yield monotonically approaches the value $(dN/do)_R$ characteristic of a randomly oriented crystal (amorphous target). For electrons, this dependence is not monotonous and has a maximum, which position depends on the observation direction. The existence of this maximum is explained by the influence of the dechanneling process on the electron motion. Generally, the obtained results demonstrate that the processes of K-shell ionization and CXR emission are rather sensitive to the properties of particle motion in oriented crystals. This opens up possibilities for applying these processes to investigate various aspects of this motion, particularly the dechanneling process.

\begin{acknowledgments}

The work was partially supported by the project No. 0124U002155 of the National Academy of Sciences of Ukraine. The work was also partially supported by the project No. 531314364 of the German Research Foundation (STCU project No. P811). 

\end{acknowledgments}

\bibliography{apssamp}

\end{document}